\documentclass[12pt]{article}

\usepackage{amsmath,amssymb}
\usepackage{array}
\usepackage{graphicx}
\usepackage{graphicx}
\usepackage{wrapfig}
\usepackage{color}
\usepackage{tensor}
\usepackage{hyperref}
\usepackage[sort,compress]{cite}
\usepackage[T1]{fontenc}
\usepackage{braket}

\numberwithin{equation}{section}

\renewcommand{\d}{\mathrm{d}}
\newcommand{\B}{\mathrm{B}}

\renewcommand{\H}{\mathrm{H}}

\renewcommand{\S}{\mathcal{S}}

\newcommand{\ent}{\mathrm{ent}}
\newcommand{\gen}{\mathrm{gen}}
\newcommand{\grav}{\mathrm{grav}}
\newcommand{\isl}{\mathrm{isl}}
\newcommand{\ext}{\mathrm{ext}}

\title{\textbf{Notes on entropy}}
\date{}
\author{}

\begin{document}
\title{\textbf{
Islands for black holes in a hybrid quantum state}}
\vspace{0.5cm}
\author{ \textbf{ Yohan Potaux$^{1}$, Debajyoti Sarkar$^{2}$ and Sergey N. Solodukhin$^{1,3}$ }} %\copyright

\date{}
\maketitle
\begin{center}
    \emph{$^{1}$Institut Denis Poisson UMR 7013,
  Universit\'e de Tours,}\\
  \emph{Parc de Grandmont, 37200 Tours, France} \\
\vspace{0.4cm}
\emph{$^{2}$Department of Physics}\\
  \emph{Indian Institute of Technology Indore}\\
  \emph{Khandwa Road 453552 Indore, India}\\
  \vspace{0.4cm}
    \emph{$^{3}$Institute for Theoretical and Mathematical Physics, } \\
\emph{Lomonosov Moscow State University, 119991 Moscow, Russia}
\end{center}

%{\vspace{-11cm}
%\begin{flushright}
%preprint
%\end{flushright}
%\vspace{10cm}
%}

%\hfill{\tt IUB-TH/***}\\\mbox{} \\
%@twocolumnfalse\endcsname

%\maketitle \thispagestyle{empty}% \vspace*{.5cm}

\vspace{0.2mm}

\begin{abstract}
\noindent{Following our previous work on hybrid quantum states in the RST model, we study its most interesting solution representing a completely regular spacetime with the structure of causal diamond,
containing an apparent horizon and radiation at infinity. Adapting recent computations of radiation entropy in terms of the entropy of entanglement, we find that this entropy follows a Page curve. This confirms our previous result \cite{Potaux_2}, which was obtained by directly calculating the thermodynamic entropy of radiation at infinity. We also investigate the presence of a possible island in these systems, and find that it does not seem to play a role in contributing to the generalized black hole entropy.}
\end{abstract}

%\noindent {PACS: 04.70Dy, 04.60.Kz, 11.25.Hf }}

\rule{7.7 cm}{.5 pt}\\
\noindent ~~~ {\footnotesize   yohan.potaux@univ-tours.fr, dsarkar@iiti.ac.in, sergey.solodukhin@univ-tours.fr}

\newpage
\tableofcontents
\pagebreak

\section{Introduction}
Ever since Hawking discovered in the 1970s that black holes emit radiation and should eventually evaporate, there has been much debate about whether their evolution is in disagreement with the principles of quantum mechanics. Indeed, according to Hawking's computation their radiation is thermal and does not correspond to a pure state, so that if a black hole (BH) is formed from a pure state its evolution would be manifestly non-unitary. This problem is known as the information paradox, or information loss problem, and many resolutions were proposed throughout the last decades to explain how information could be preserved. For a nice introduction to this problem and a presentation of possible solutions one can consult \cite{Harlow_Lectures}. In fact many people were convinced that black hole evolution should indeed be unitary when in 1997 Maldacena \cite{Maldacena_1997} exhibited the correspondence between Anti de Sitter (AdS) quantum gravity and conformal field theory (CFT), a realization of the holographic principle proposed by 't Hooft and Susskind in 1993 \cite{tHooft:1993dmi,Susskind_WorldHologram}. The basic idea is that, if one considers a black hole in an AdS spacetime, the equivalent CFT on the boundary evolves unitarily, so the black hole should do too. However this is not a consensus, see for instance \cite{Mathur_InfoParadoxIntro} for a discussion about the relevance of this argument.

\begin{figure}[htb]
 \centering
 \includegraphics[scale=0.8]{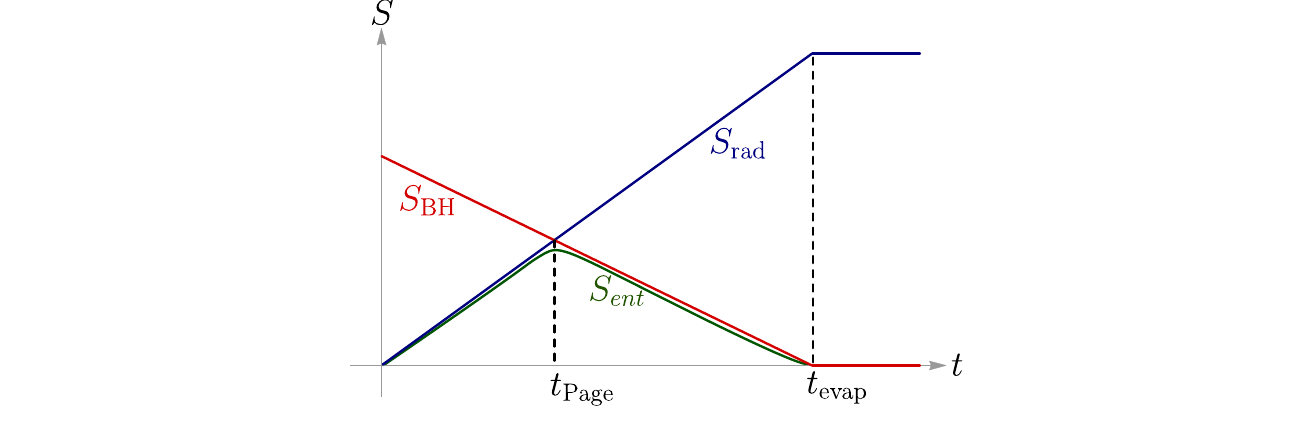}
 \caption{Schematic plot of the various entropies during a black hole evaporation process. The entanglement entropy of the radiation $S_\ent$, also known as the ``fine-grained'' entropy, is always smaller than the thermodynamic, or ``coarse-grained'' entropy of the black hole $S_{\B\H}$, and the radiation entropy $S_\mathrm{rad}$. If the evolution is unitary, $S_\ent$ has to start and end at zero, and it is expected that its maximum should coincide with the point where $S_{\B\H} = S_\mathrm{rad}$.}
 \label{FIG_Intro_PageCurve}
\end{figure}

A criteria was suggested by Page \cite{Page_1993,Page_2013} to establish whether an evaporation process is unitary and preserves information. The idea is to consider the evolution of the Hawking radiation's entanglement entropy during a unitary evaporation process. Initially, when the black hole has just formed, there is no radiation and this entropy is zero. As radiation starts coming out, it is entangled with the inside of the black hole, so even though the whole quantum state remains pure (as evolution is unitary), the subsystem consisting of this outside radiation has a positive entanglement entropy. When the black hole has completely evaporated, there remains only outside radiation and it has nothing left to be entangled with, so the entropy must be back to zero. What this means is that there exists a time at which the entanglement entropy of the Hawking radiation starts decreasing, signaling that trans-horizon correlations start to appear in the radiation. This time is referred to as the Page time and should be of order half the total evaporation time \cite{Page_2013}. Plotting the evolution of the entropy of radiation as a function of time, we thus get a curve that is increasing at first and then decreasing, called the Page curve. A schematic plot is presented on figure \ref{FIG_Intro_PageCurve}.

In the last decade, a new procedure was introduced with the aim of computing the entanglement entropy during an evaporation process. The idea is to introduce so-called islands, as part of the quantum extremal surface \cite{Engelhardt:2014gca}, which are surfaces located inside the horizon (and sometimes outside) that contribute to the generalized entropy. See \cite{Almheiri_EntropyHawkingRadiation} for a nice introduction to this procedure, which was used in \cite{Almheiri_Engelhardt} to derive a Page curve for Jackiw-Teitelboim gravity, a model of two-dimensional semi-classical gravity.

In this paper we pursue the work done in \cite{Potaux_1,Potaux_2,Potaux_3}, in which we focused on another model of two-dimensional semi-classical gravity, the RST model \cite{RST_0}. This model is a semi-classical extension of the classical CGHS model \cite{CGHS_Original}, obtained by considering conformally coupled quantum fields, whose energy-momentum tensor is fixed by the conformal anomaly in two dimensions. The RST action is given by
\begin{equation}
 \S = \frac{1}{2\pi}\int\d^2x\,\sqrt{-g}\,
 \Big\{e^{-2\phi}\big[R + 4(\nabla\phi)^2+4\lambda^2\big]
 -\frac{1}{2}(\nabla f)^2
 -\kappa\phi R
 -\frac{\kappa}{2}R\,\square^{-1}R
 \Big\}
 \,,
\end{equation}
with $\phi$ the dilaton field and where $\kappa$ is proportional to the central charge of the conformal field theory, \textit{i.e.} to the number of fields if we are only considering scalar fields. $\lambda$ is the cosmological constant in the theory, and $f$ will denote infalling classical matter, whose presence will turn a static solution into dynamical one.

In our previous papers, especially \cite{Potaux_3}, we were able to define the quantum states of interest, namely the Hartle-Hawking, Boulware and Unruh states, and find the corresponding solutions in the RST model. An interesting scenario is obtained when considering fields in the Boulware state, meaning that they do not radiate at infinity in a static spacetime. The Minkowski vacuum is one particular instance of this case. When sending in a shockwave, an apparent horizon and a singularity are created. An interesting point is that, although the fields are initially in the Boulware state, the creation of a horizon effectively causes them to radiate at future infinity. At first, the singularity is hidden behind the horizon, but they collide at a finite point, signaling the breakdown of the semi-classical approximation and the evaporation of the black hole. In \cite{Hartman_Islands,Gautason:2020}, based on the work of \cite{Fiola_BHThermo}, the authors studied this solution in the context of the information loss problem. First, they computed the entanglement entropy without considering islands, finding that it increases up until the evaporation point, indicating a non-unitary evolution. However, by applying the island procedure, they found that one eventually obtains a Page curve like behavior. This provides a crucial example, where considerations of islands can indeed signal a unitary evolution of the BH.

In our previous work, we provided a systematic and detailed study of these various states, uncovered a new set of states, and explored the possibility of combining two different states creating a hybrid quantum state. In the last case, the idea is to explore what happens if the backreacting conformal fields are non-physical (such as ghosts), and how the physics changes when they are considered in combination with the usual physical fields. These fields contribute negatively to the total central charge \cite{Russo:1992yg,deAlwis:1992hv,Bilal:1992kv} and we argued that the most natural state for them is the Boulware state, as they should not be observed at infinity. As for the physical fields, a natural choice is the Unruh state,\footnote{One can also take the physical fields to be in the Hartle-Hawking state, which corresponds to a black hole in thermal equilibrium with its radiation. The only notable difference between the two states is the presence of incoming radiation at past infinity.} which contains outgoing radiation at future infinity but no incoming radiation at past infinity. Thus it is expected to be the most suited state to describe black hole evaporation. In this setup, the most interesting case arises when the total central charge is negative, $\kappa < 0$, that is when the non-physical fields dominate the physical ones. Indeed, the resulting solution is singularity free, asymptotically flat, and sending in a shock wave creates an apparent horizon but no singularity. Importantly, the radiation at future infinity has two contribution: one from the physical fields, which is thermal, and one from the non-physical fields, which can be interpreted as corrections rendering the total radiation non-thermal. In fact, we argued in \cite{Potaux_2,Potaux_3} that the thermodynamic entropy of the resulting radiation followed a Page curve thanks to the radiation of non-physical particles at future infinity.

What we would like to do now is study the entanglement entropy of the radiation of this hybrid solution, with and without islands. In order to do this, we will adapt the discussion from \cite{Hartman_Islands,Gautason:2020} on the Boulware state solution, to our solution. We want to investigate whether the inclusion of islands is relevant in this case, and what are its implications towards the Page curve with or without its contribution.

In section \ref{sec:setup}, we briefly review how the entanglement entropy can be defined, and reproduce the main results of \cite{Hartman_Islands,Gautason:2020} in section \ref{Section_EntanglementEntropyBoulware}. In section \ref{Section_EntanglementEntropyUnruh} we study how the same computation translates to our hybrid solution. We find that one indeed obtains a Page curve even without an island. Finally, section \ref{sec:island} is dedicated to the island procedure, where we once again follow \cite{Hartman_Islands,Gautason:2020} to investigate the relevance of islands for our hybrid case. Sections \ref{Section_EntanglementEntropyUnruh} and \ref{sec:island} present the main new results of our paper. We conclude in section \ref{sec:conclude}.

%We will work exclusively in the conformal gauge, so that the metric can be expressed as
%\begin{equation}
% \d s^2 = -e^{2\rho}\d \sigma^+ \d \sigma^-
% \,.
%\end{equation}

%%%%%%%%%%%%%%%%%%%%%%%%%%%%%%%%%%%%%%%%%%%%%%%%%%%
\section{Definition of the entanglement entropy}\label{sec:setup}
%%%%%%%%%%%%%%%%%%%%%%%%%%%%%%%%%%%%%%%%%%%%%%%%%%%

To define the entanglement entropy, we follow the paper by Hartman et al.~\cite{Hartman_Islands}, which is based on an earlier work by Fiola et al.~\cite{Fiola_BHThermo}. We will work exclusively in the conformal gauge, so that the metric can be expressed as
\begin{equation}
 \d s^2 = -e^{2\rho}\d \sigma^+ \d \sigma^-
 \,,
\end{equation}
with $\rho$ the conformal factor.

\begin{figure}[htb]
 \centering
 \includegraphics[scale=0.7]{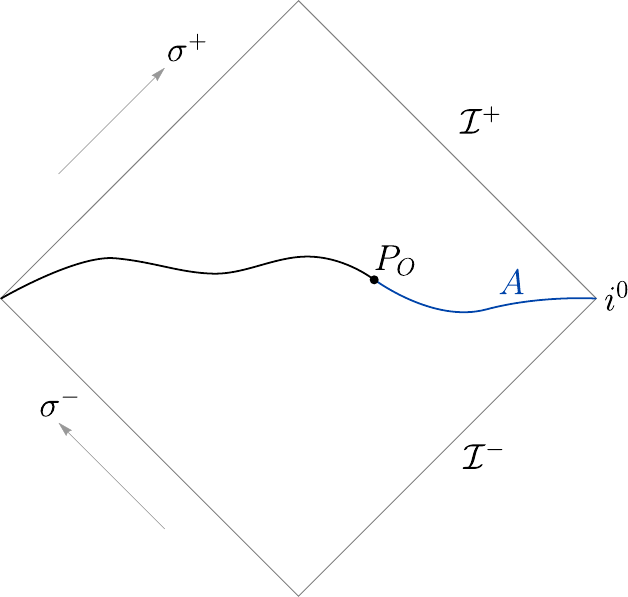}
 \caption{An everywhere spacelike or null slice in a causal diamond. On the whole slice the state of the quantum fields is pure, but when considering a portion $A$ of this slice it becomes mixed and there is an associated entanglement entropy.}
 \label{FIG_Slice}
\end{figure}

We will always consider the fields to be in their vacuum state at past infinity $\mathcal{I}^-$, where the vacuum state is defined with respect to the corresponding asymptotically flat coordinates $(\sigma^+,\sigma^-)$. This is physically required by the fact that there should be no incoming radiation, and it can correspond to either the Boulware or Unruh states. Let us consider a causal diamond and a full slice, everywhere spacelike or null, going from spatial infinity $i^0$ to the other end of the diamond, as shown on figure \ref{FIG_Slice}.
On such a slice, the quantum state of the fields is pure and has zero entanglement entropy. However, it is a mixed state on a portion $A$ of this slice, going from $i^0$ to some interior point $P_O = (\sigma^+_O,\sigma^-_O)$, which can be thought of as the location of an observer. For $N$ scalar fields, the corresponding entanglement entropy is given by \cite{Fiola_BHThermo,Hartman_Islands},
\begin{equation}
 S_{\mathrm{ent}}(A) = \frac{N}{6}\Big(\rho_O + \ln\frac{L}{\epsilon}\Big)
 \,,
\end{equation}
where $\rho_O \equiv \rho(P_O)$ is the conformal factor in the $(\sigma^+,\sigma^-)$ coordinates at point $P_O$, and $L$ and $\epsilon$ are IR and UV regulators respectively. Using $\kappa = \frac{N}{24}$ we get that
\begin{equation}
 S_\ent = 4\kappa\Big(\rho_O + \ln\frac{L}{\epsilon}\Big)
 \,.
\end{equation}
Let us now study this quantity for the Boulware and hybrid solutions.

%%%%%%%%%%%%%%%%%%%%%%%%%%%%%%%%%%%%%%%%%%%%%%%%%%%%%%
\section{Entanglement entropy for the Boulware state}
\label{Section_EntanglementEntropyBoulware}
%%%%%%%%%%%%%%%%%%%%%%%%%%%%%%%%%%%%%%%%%%%%%%%%%%%%%%

\begin{figure}[htb]
 \centering
 \includegraphics[scale=0.7]{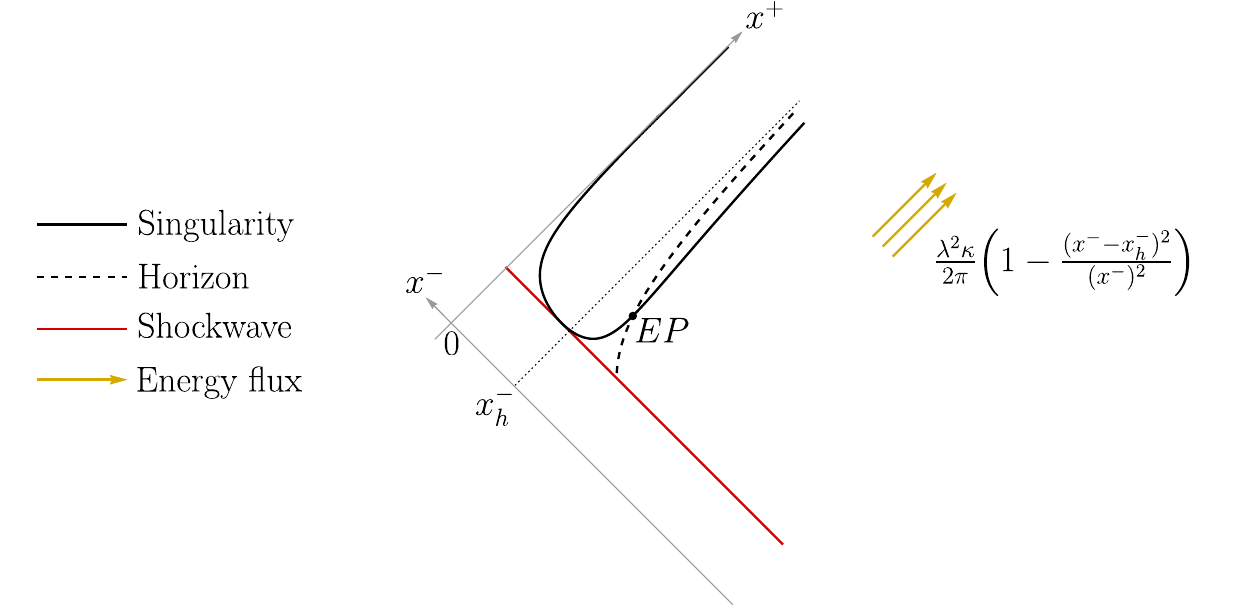}
 \caption{Spacetime for the solution describing physical particles in the Boulware state. At first it is Minkowski, but after the shockwave an apparent horizon and a singularity appear and collide at a finite point $EP$, interpreted as the evaporation point.}
 \label{FIG_BoulwarePhysicalforEntropy}
\end{figure}

We will start by computing the entanglement entropy for the solution describing physical fields in the Boulware state. We therefore assume that $\kappa > 0$ in this section. For convenience, let us recall its important properties (see \cite{Potaux_3} for a detailed presentation). The metric is $\d s^2 = -e^{2\phi} \d x^+ \d x^-$, and the value of the dilaton $\phi (x^+,x^-)$ is determined by the master equation
\begin{equation}
 \Omega = e^{-2\phi} + \kappa\phi
 = -\lambda^2 x^+x^- - \frac{\kappa}{2}\ln(-\lambda^2x^+x^-) + \frac{m}{\lambda}\Big(1-\frac{x^+}{x^+_0}\Big)\theta(x^+-x^+_0)
 \,.
 \label{EQ_BoulwarePhysical_PerturbedSolution}
\end{equation}
The parameter $m$ is the mass contained in the incoming pulse of classical matter along $x^+ = x^+_0$. Before the shockwave ($x^+<x^+_0$), this spacetime is the Minkowski vacuum, and after the shock ($x^+<x^+_0$), a singularity appears on the curve defined by
\begin{equation}
 f(W) \equiv W - \ln W =
 1
 +\frac{2m}{\lambda\kappa}\Big(\frac{x^+}{x^+_0}-1\Big)
 \,, \quad
 W \equiv -\frac{2}{\kappa}\lambda^2x^+x^-
 \,.
\end{equation}
An apparent horizon also appears on the curve\footnote{There is actually another horizon, but it is located behind the singularity at all times and is therefore not relevant \cite{Potaux_3} .}
\begin{equation}
 -\lambda^2x^+(x^- - x^-_h) = \frac{\kappa}{2}
 \,, \quad
 x^-_h \equiv -\frac{m}{\lambda^3x^+_0}
 \,.
\end{equation}
This horizon intersects the singularity at a finite point, interpreted as the endpoint of the evaporation in \cite{Fiola_BHThermo,Hartman_Islands}, and whose coordinates $(x^+_{EP},x^-_{EP})$ are given by
\begin{equation}
 x^+_{EP} = \frac{\lambda\kappa}{2m}(e^{\frac{2m}{\lambda\kappa}}-1)\,x^+_0 > x^+_0
 \,,
 \quad
 x^-_{EP} = \frac{x^-_h}{1-e^{-\frac{2m}{\lambda\kappa}}}
 < x^-_h
 \,.
\end{equation}
This spacetime is represented on figure \ref{FIG_BoulwarePhysicalforEntropy}.

At past infinity, that is in the limit $x^- \rightarrow -\infty$, we have $e^{-2\phi} \sim -\lambda^2x^+x^-$, so the spacetime is asymptotically flat as can be seen by using the asymptotically flat coordinates $(\sigma^+,\sigma^-)$ defined by 
\begin{equation}
 \sigma^\pm = \pm\frac{1}{\lambda}\ln(\pm \lambda x^\pm)
 \,, \quad
 x^\pm = \pm\frac{1}{\lambda}e^{\pm\lambda \sigma^\pm}
 \,.
 \label{EQ_AsymptoticallyFlatCoordinates}
\end{equation}
In these coordinates, the metric becomes $\d s^2 = -e^{2\rho}\d\sigma^+\d\sigma^-$, with the conformal factor
\begin{equation}
 \rho = \phi + \frac{1}{2}\ln(-\lambda^2x^+x^-)
 \,.
\end{equation}
As explained in the previous section, this quantity is directly related to the entanglement entropy of a slice between spatial infinity and the point $(x^+_O,x^-_O)$ according to
\begin{equation}
 S_\ent(x^+_O,x^-_O) = 4\kappa\Big(\rho(x^+_O,x^-_O) + \ln\frac{L}{\epsilon}\Big)
 \,.
\end{equation}
The regulators are there to regularize the divergences but don't modify the global behavior of the entropy and cancel out when considering a change in entropy. Note that for the Minkowski vacuum, $\rho$ is zero everywhere and does not contribute to the entanglement entropy. What we can do to study this entropy is to fix some value of $x^+_O$ and see how it evolves as a function of the retarded time $x^-_O$. More precisely, let us define
\begin{equation}
 \Delta S_\ent(x^+_O,x^-_O) \equiv S_\ent(x^+_O,x^-_O) - S_\ent(x^+_O,-\infty) = 4\kappa\rho(x^+_O,x^-_O)
 \,,
\end{equation}
since $\rho \rightarrow 0$ at past infinity ($x^- \rightarrow -\infty$) by definition. Before the shock, this quantity is zero everywhere, so we will focus on what happens after the shock ($x^+_O>x^+_0$) in the following. Let us then determine the sign of $\partial_-\rho$. Denoting $\Omega' \equiv \frac{\d\Omega}{\d\phi} = \kappa - 2e^{-2\phi}$, we have 
\begin{align}
 \partial_-\rho & = \partial_-\phi + \frac{1}{2x^-}
 \,, \nonumber
 \\
 & = \frac{1}{\Omega'}\partial_-\Omega + \frac{1}{2x^-}
 \,, \nonumber
 \\
 & = -\frac{\lambda^2x^+}{\Omega'}(1-e^{-2\rho})
 \,.
\end{align}
As $\Omega' < 0$ outside of the singularity, which is defined by $\Omega'=0$, and $x^+ >0$ on the right quadrant, the sign of $\partial_-\rho$, and therefore of $\partial_-\Delta S_\ent$, is determined by the sign of $\rho$. Concretely, the behavior of the change in entropy $\Delta S_\ent$ will be fully determined by the sign of $\rho$ in the limit $x^- \rightarrow -\infty$. If $\rho$ is initially positive (respectively negative) there, it will grow (respectively decrease) indefinitely, or at least until the evaporation point.

\begin{figure}[htb]
 \centering
 \includegraphics[scale=0.7]{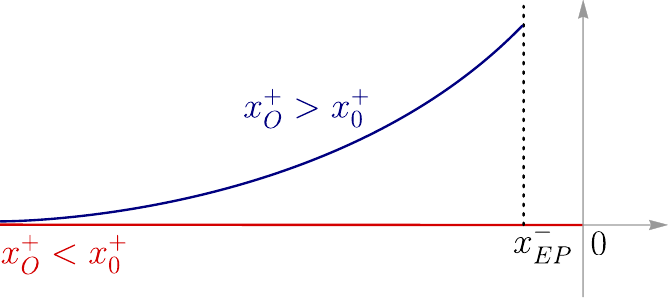}
 \caption{Evolution of the entanglement entropy along a null line $x^+=x^+_O$ for the spacetime describing physical fields in the Boulware state, perturbed by a shockwave at $x^+=x^+_0$. Before the shock, spacetime is Minkowski and the entanglement entropy is constant, so that $\Delta S_\ent = 0$. However, after the shock, it grows as a function of the retarded time, suggesting a non-unitary evolution.}
 \label{FIG_Entropy_BoulwarePhysical}
\end{figure}

In this limit, $e^{-2\phi} = -\lambda^2x^+x^-e^{-2\rho} \sim -\lambda^2x^+x^-$ so that $\rho \rightarrow 0$, but we still have to determine whether it approaches $0$ from above or from below. Plugging $\phi = -\frac{1}{2}\ln(-\lambda^2x^+x^-) + \rho$ in \eqref{EQ_BoulwarePhysical_PerturbedSolution}, we get that, after the shock,
\begin{equation}
 \kappa\rho - \lambda^2x^+x^-e^{-2\rho} =
 -\lambda^2x^+x^- + \frac{m}{\lambda}\Big(1-\frac{x^+}{x^+_0}\Big)
 \,.
\end{equation}
Using the fact that $\rho \rightarrow 0$ at past infinity, we obtain that
\begin{equation}
 \lambda^2x^+x^- (1-e^{-2\rho}) \sim
 2\lambda^2x^+x^-\rho
 \rightarrow
 \frac{m}{\lambda}\Big(1-\frac{x^+}{x^+_0}\Big)
 \,,
\end{equation}
so that
\begin{equation}
 \rho \sim \frac{1}{2\lambda^2x^+x^-}
 \frac{m}{\lambda}\Big(1-\frac{x^+}{x^+_0}\Big) > 0
 \,.
 \label{EQ_Behavior_Rho_Boulware_PastInfinity}
\end{equation}
Therefore $\rho \rightarrow 0^+$ for $x^+ > x^+_0$. This implies that the change in entropy $\Delta S_\ent$ grows as the retarded time increases, reaching a finite positive value at $x^-_{EP}$. This suggests that evolution is not unitary and that information is destroyed. This is represented on figure \ref{FIG_Entropy_BoulwarePhysical}.

It is also interesting to study the behavior of the entropy when the observer $P_O$ is at future infinity $\mathcal{I}^+$, i.e. as a function of $x^-_O$ in the limit $x^+_O \rightarrow +\infty$. Then $\Omega_O \sim e^{-2\phi_O} \sim -\lambda^2x^+_O(x^-_O-x^-_h)$, which leads to
\begin{equation}
 \Delta S_\ent(x^+_O,x^-_O) \rightarrow \Delta S_\ent^\infty (x^-_O)
 \equiv -2\kappa \ln\Big(1-\frac{x^-_h}{x^-_O}\Big)
 = 2\kappa \ln\Big(1+\frac{x^-_h}{x^-_O-x^-_h}\Big)
 \,.
 \label{EQ_Boulware_Physical_Entropy}
\end{equation}
The asymptotically flat coordinates at $\mathcal{I}^+$ are $(\tilde{\sigma}^+,\tilde{\sigma}^-)$, defined by
\begin{equation}
 \lambda x^+ = e^{\lambda\tilde{\sigma}^+}
 \,, \quad
 \lambda(x^- - x^-_h) = -e^{-\lambda\tilde{\sigma}^-}
 \,,
\end{equation}
so that, as a function of the affine retarded time $\tilde{\sigma}^-_O$, we have
\begin{equation}
 \Delta S_\ent^\infty(\tilde{\sigma}^-_O) = 
 2\kappa\ln(1-\lambda x^-_h e^{\lambda\tilde{\sigma}^-_O})
 \,.
 \label{EQ_Boulware_Physical_DeltaS_TildeSigmaO}
\end{equation} 
Finally, at the evaporation endpoint $EP$ we get
\begin{equation}
 \Delta S_\ent^\infty (x^-_{EP}) = \frac{4m}{\lambda}
 \,.
\end{equation}
This is twice the entropy of the classical black hole of mass $m$. These results agree with equations (2.39) and (2.40) from \cite{Hartman_Islands}.

%%%%%%%%%%%%%%%%%%%%%%%%%%%%%%%%%%%%%%%%%%%%%%%%%%%%%
\section{Entanglement for the Unruh/Boulware state}
\label{Section_EntanglementEntropyUnruh}
\begin{figure}[htb]
 \centering
 \includegraphics[scale=0.7]{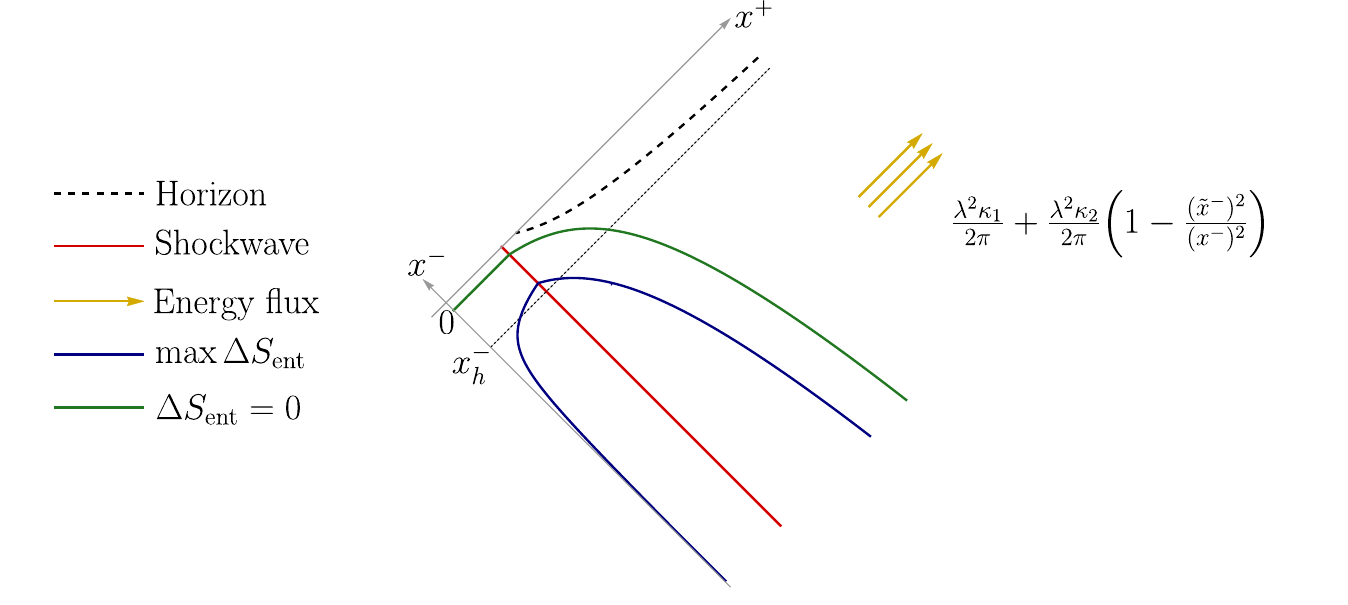}
 \caption{Spacetime for the hybrid Boulware-Unruh solution, an apparent horizon forms after the shock wave of incoming matter. The blue curve corresponds to the location of the maximum of the entanglement entropy $S_\ent$ along a null line of constant $x^+$, and the green curve is where the change in entropy vanishes.}
 \label{FIG_UnruhBoulwareSpacetime}
\end{figure}

%%%%%%%%%%%%%%%%%%%%%%%%%%%%%%%%%%%%%%%%%%%%%%%%%%%%%
Now we will try to apply the same procedure to the hybrid solution, describing physical fields in the Unruh state and non-physical fields in the Boulware state (see \cite{Potaux_3} for a detailed study of such states). The corresponding master equation is
\begin{equation}
 \Omega = e^{-2\phi} + \kappa\phi
 = -\lambda^2x^+x^-
 - \frac{\kappa_1}{2}\ln(\lambda x^+)
 - \frac{\kappa_2}{2}\ln(-\lambda^2x^+x^-)
 + \frac{M}{\lambda}
 + \frac{m}{\lambda}\Big(1-\frac{x^+}{x^+_0}\Big)
 \theta(x^+-x^+_0)
 \,,
 \label{EQ_UnruhMasterEquation}
\end{equation}
where $\kappa_1 > 0$ is the central charge contribution of the physical fields, and $\kappa_2 < 0$ that of the non-physical fields. The total central charge $\kappa = \kappa_1 + \kappa_2$ is taken to be negative, $\kappa < 0$.
Even for $M = 0$, the spacetime before the shock is not Minkowski, although it is geodesically complete and singularity free. After the shock, an apparent horizon forms along the curve defined by
\begin{equation}
 -\lambda^2x^+(x^--x^-_h) = \frac{\kappa}{2}
 \,,
\end{equation}
and an observer at future infinity will observe thermal radiation from physical particles at temperature $T = \frac{\lambda}{2\pi}$. As discussed in \cite{Potaux_2,Potaux_3}, there is also a non-thermal flux of non-physical particles, and we earlier showed that it leads to a Page curve for the entropy of radiation, by directly computing the asymptotic radiation. However, one might ask whether the same can be shown by computing the evolution of the entanglement entropy. Note that, strictly speaking, there is no evaporation and the apparent horizon goes all the way to null future infinity at the point $\mathcal{C} = (+\infty, x^-_h)$. This spacetime is represented on figure \ref{FIG_UnruhBoulwareSpacetime}.

Just as for the Boulware solution, we can study the entanglement entropy, which is still given by $S_\ent = 4\kappa\big(\rho + \ln\frac{L}{\epsilon}\big)$ with
\begin{equation}
 \rho = \phi + \frac{1}{2}\ln(-\lambda^2x^+x^-)
 \,.
\end{equation}
This is because, at past infinity the flat coordinates are still given by \eqref{EQ_AsymptoticallyFlatCoordinates}. Note that, as $\kappa$ is negative, $\Delta S_\ent = 4\kappa\rho$ has the opposite sign compared to $\rho$. Let us study its behavior along a null line $x^+ = \mathrm{constant}$.

First, when $x^- \rightarrow -\infty$, we can follow a similar reasoning to the Boulware case and obtain
\begin{equation}
 \rho \sim \frac{1}{2\lambda^2x^+x^-}
 \biggl[\kappa_1\ln(-\lambda x^-) + \frac{M}{\lambda}
 + \frac{m}{\lambda}\Big(1-\frac{x^+}{x^+_0}\theta(x^+-x^+_0)\Big)\bigg]
 \rightarrow 0^-
 \,.
\end{equation}
Note the difference with \eqref{EQ_Behavior_Rho_Boulware_PastInfinity}, the extra term proportional to $\kappa_1$ ensures that $\rho \rightarrow 0^-$ for any value of $x^+$, and therefore that $\Delta S_\ent \rightarrow 0^+$.

When $x^- \rightarrow 0$ we have
\begin{equation}
 \Omega \sim \kappa\phi \sim -\frac{\kappa_2}{2}\ln(-\lambda x^-)
 \,,
\end{equation}
so that $\rho \sim \frac{\kappa_1}{2\kappa}\ln(-\lambda x^-) \rightarrow +\infty$. More precisely,
\begin{equation}
 \rho \sim \frac{\kappa_1}{2\kappa}\ln(-\lambda x^-) + \frac{M}{\lambda\kappa} + \frac{m}{\lambda\kappa}\biggl(1-\frac{x^+}{x^+_0}\biggr)\theta(x^+-x^+_0)
 \,,
\end{equation}
so that
\begin{equation}
 \Delta S_\ent \sim 2\kappa_1\ln(-\lambda x^-) + \frac{4M}{\lambda} + \frac{4m}{\lambda}\biggl(1-\frac{x^+}{x^+_0}\biggr)\theta(x^+-x^+_0)
 \rightarrow - \infty
 \,.
\end{equation}
These two limits suggest that the change in entropy $\Delta S_\ent$ reaches a maximum in the bulk and this is indeed the case. The condition $\partial_-\rho = 0$ coupled to the master equation \eqref{EQ_UnruhMasterEquation} leads to the equation
\begin{equation}
% \Delta S_\ent \bigg|_{\partial_-\Delta S_\ent = 0} = 4\kappa\rho\bigg|_{\partial_-\rho = 0} =
 -2\kappa \ln\frac{-\lambda^2x^+x^- + \frac{\kappa_1}{2}}{-\lambda^2x^+x^-} = 2\kappa_1\ln(-\lambda x^-) - 2\kappa_1 + \frac{4M}{\lambda} + \frac{4m}{\lambda}\Big(1-\frac{x^+}{x^+_0}\Big)\theta(x^+-x^+_0)
 \,,
\end{equation}
which defines a curve contained in the spacetime and represented on figure \ref{FIG_UnruhBoulwareSpacetime}. Thus the entanglement entropy along a null line of constant $x^+$ exhibits the typical behavior of the Page curve: it grows up until a maximum and then decreases. It vanishes on the curve where $\rho=0$ which is simply given by
\begin{equation}
 0 = \kappa_1\ln(-\lambda x^-) + \frac{2M}{\lambda} +
 \frac{2m}{\lambda}\Big(1-\frac{x^+}{x^+_0}\Big)\theta(x^+-x^+_0)
 \,,
\end{equation}
and also represented on figure \ref{FIG_UnruhBoulwareSpacetime}. Note that before the shock, this curve corresponds to a constant value of $x^-$.

\begin{figure}[htb]
 \centering
 \includegraphics[scale=0.7]{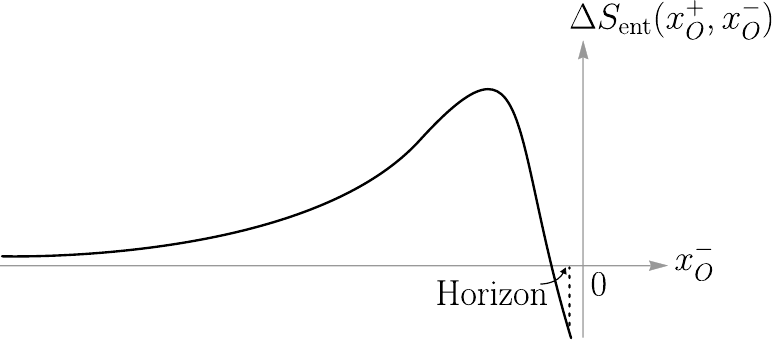}
 \caption{Schematic plot of the change in entanglement entropy along an incoming null line $x^+=\mathrm{constant}$ for the Hybrid Unruh/Boulware solution, which shows the typical behavior for the Page Curve. Note that the location of the maximum depends on the value of $x^+$, as shown on figure \ref{FIG_UnruhBoulwareSpacetime}.}
 \label{FIG_Entropy_UnruhBoulware}
\end{figure}

It appears that for any fixed value of $x^+$, the entanglement entropy follows a Page curve as a function of the retarded time $x^-$. The behavior of the entanglement entropy along an incoming null line is plotted schematically on figure \ref{FIG_Entropy_UnruhBoulware}. As $x^+$ increases, that is when getting closer to null infinity, the maximum moves towards earlier retarded times.
%On the horizon the entropy reaches a finite value $\Delta S_\ent^h(x^+) = 4\kappa\rho_h(x^+)$ where $\rho_h(x^+)$ is determined by solving the equation
%\begin{equation}
% \kappa\rho_h + \biggl(\frac{\kappa}{2} + \frac{mx^+}{\lambda x^+_0}\biggr)e^{-2\rho_h} = \frac{\kappa}{2}
% + \frac{\kappa_1}{2}\ln\biggl( \frac{\kappa}{2\lambda x^+} + x^-_h \biggr) + \frac{M+m}{\lambda}
% \,.
%\end{equation}
%In the limit $x^+ \rightarrow +\infty$ the only possibility is $\rho_h \rightarrow +\infty$ \textit{i.e.} $\Delta S_\ent^h \rightarrow -\infty$.

At future infinity ($x^+ \rightarrow \infty$) the entanglement entropy goes as
\begin{equation}
 \Delta S_\ent \rightarrow \Delta S_\ent^\infty(x^-) =
 -2\kappa\ln\biggl(1-\frac{x^-_h}{x^-}\biggr)
 \,,
\end{equation}
which is actually the same as for the Boulware case (see \eqref{EQ_Boulware_Physical_Entropy}), but here $\kappa$ is negative. This is a monotonically decreasing function of the retarded time $x^-$, which makes sense as when $x^+ \rightarrow \infty$, the maximum of $\Delta S_\ent$ is sent to past infinity, as seen by looking at the blue curve on figure \ref{FIG_UnruhBoulwareSpacetime}. However, for an observer close to future infinity, but at a finite point, this maximum is also located at a finite point and he does observe a Page curve behavior for the change in entanglement entropy $\Delta S_\ent$.

%%%%%%%%%%%%%%%%%%%%%%%%%%%%%%%%%%
\section{Islands}\label{sec:island}
%%%%%%%%%%%%%%%%%%%%%%%%%%%%%%%%%%

We will now discuss how the island procedure affects the results obtained in the previous two sections. We will start by briefly presenting the idea behind islands, and then we will closely follow the computations of \cite{Hartman_Islands} to see how a Page curve can be derived for the Boulware solution (even though the analysis from section \ref{Section_EntanglementEntropyBoulware} suggested that evolution was not unitary). Finally we will try to apply it to our hybrid Unruh/Boulware solution and see if it changes the result from last section, where a Page curve already seemed to be present.

\begin{figure}[htb]
 \centering
 \includegraphics[scale=0.7]{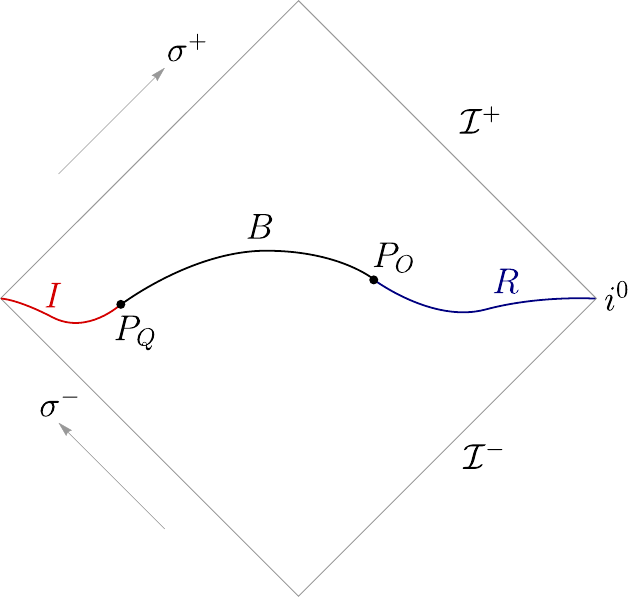}
 \caption{A full Cauchy surface that is everywhere spacelike or null, separated in three parts. Region $I$ is the island, with $P_Q$ the quantum extremal surface obtained after extremization.}
 \label{FIG_Island_Slice}
\end{figure}

The island procedure associates an entropy $S_\isl(P_O)$ to an observer located at a point $P_O$ by \cite{Engelhardt:2014gca,Penington_2019,Almheiri_Engelhardt}
\begin{equation}
 S_\isl(P_O) = \min\, \underset{P_Q}{\ext}\, S_\gen(I\cup R)
 \,,
 \label{EQ_SIslandDefinition}
\end{equation}
where $C = I\cup B \cup R$ defines a complete Cauchy surface that is everywhere spacelike or null, and which contains the two points $P_O$ and $P_Q$, and $B$ is the surface between $P_Q$ and $P_O$. See figure \ref{FIG_Island_Slice} for an illustration of the surface, to be applied in various scenarios. 
The region $I$ between the left boundary and $P_Q$ is called the island, and region $R$ is where the radiation passes through, that is between right spatial infinity and point $P_O$.

What \eqref{EQ_SIslandDefinition} tells us is that we should compute the generalized entropy $S_\gen(I\cup R)$, extremize it with respect to the location of point $P_Q$, and then take the minimum between the resulting entropy and the entropy computed without any island (\textit{i.e.} the one computed previously). We could also consider that there may be more than one island, but for simplicity's sake we will restrict ourselves to the one island case.

The generalized entropy associated to a point $P_Q$ is
\begin{equation}
 S_\gen(I \cup R) = S_\grav(P_Q) + S_\ent(I\cup R)
 \,,
\end{equation}
with $S_\grav(P_Q)$ the Bekenstein-Hawking entropy of the island. In the context of the RST model it is given by (see e.g. equation (98) of \cite{Fiola_BHThermo}),
\begin{equation}
 S_\grav(P_Q) = 2e^{-2\phi_Q} - 2\kappa\phi_Q + 4\kappa\ln\lambda\epsilon
 \,, \quad
 \phi_Q \equiv \phi(P_Q)
 \,,
\end{equation}
where $\epsilon$ is an UV regulator with dimension of length.
The entanglement entropy $S_\ent(I \cup R)$ can then be obtained as (once again, see e.g. equation (3.5) of \cite{Hartman_Islands}, or equation (64) of \cite{Fiola_BHThermo}),
\begin{equation}
 S_\ent(I\cup R) = 4\kappa(\rho_Q + \rho_O) +  4\kappa\ln\frac{(\sigma^-_Q - \sigma^-_O)(\sigma^+_O - \sigma^+_Q)}{\epsilon^2}
 \,.
\end{equation}
We will work in the asymptotically flat coordinates $\sigma^\pm$ as it makes the calculations easier. Since $\rho = \phi + \frac{\lambda}{2}(\sigma^+-\sigma^-)$, we obtain (using $\Omega = e^{-2\phi} + \kappa\phi$)
\begin{equation}
 S_\gen(I\cup R) = 2\Omega_Q
 + 4\kappa \bigg[\frac{\lambda}{2}(\sigma^+_Q - \sigma^-_Q)
 + \ln\frac{(\sigma^-_Q - \sigma^-_O)(\sigma^+_O - \sigma^+_Q)}{e^{-\rho_O}\epsilon/\lambda}
 \bigg]
 \,.
 \label{EQ_Sgen_General}
\end{equation}
In order to compare to the computations without islands done in the previous two sections, we will assume that the point $P_O$ is located at future null infinity. So we will take the limit $\sigma_O^+ \rightarrow +\infty$ in \eqref{EQ_Sgen_General}, and then extremize the resulting expression with respect to $\sigma^\pm_Q$ to find the location of the quantum extremal surface $P_Q$. Finally, to get the island entropy $S_\isl(P_O)$, we have to take the minimum between the entropy computed with and without the island. We will first carry this out for the Boulware state solution, recovering the results of \cite{Hartman_Islands}. Afterwards, we will repeat our analysis for the hybrid Unruh/Boulware solution.

\subsection{Island for the Boulware solution}
\begin{figure}[htbp]
 \centering
 \includegraphics[scale=0.7]{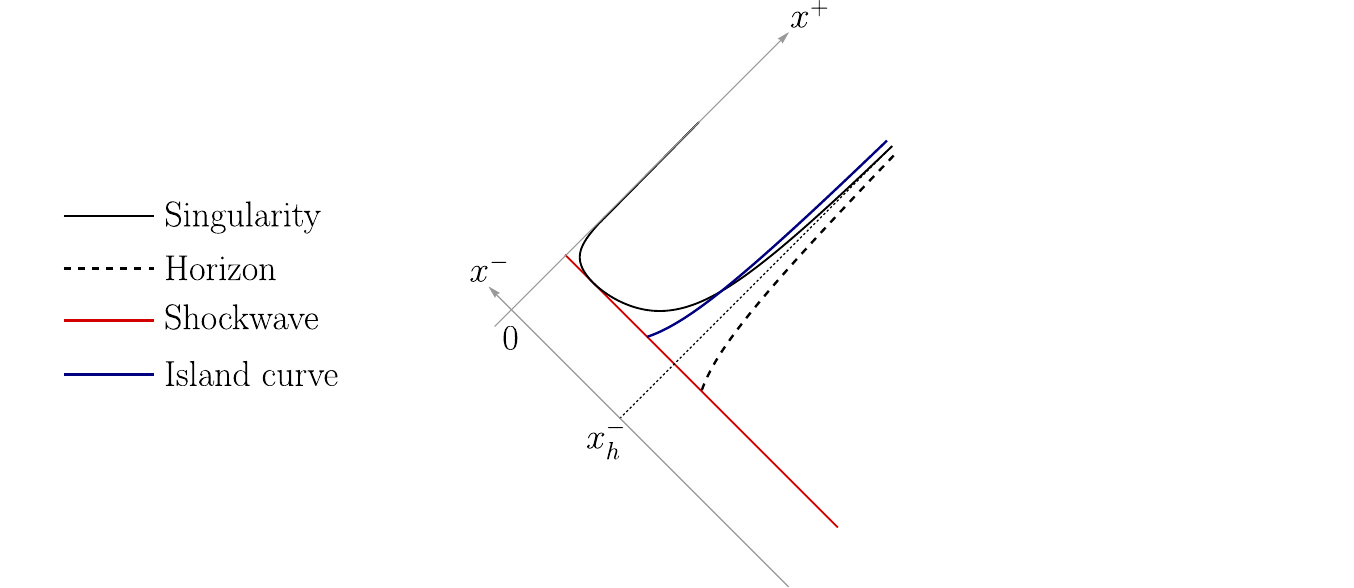}
 \caption{The so-called island curve in blue indicates where the quantum extremal surface $P_Q$ can be located (if it exists), although its location is not necessarily monotonic along this curve. This curve is partially outside the singularity if and only if $m > \lambda\kappa$. Otherwise it is completely hidden behind the singularity and there is no island.}
 \label{FIG_BoulwareIsland}
\end{figure}

According to \eqref{EQ_BoulwarePhysical_PerturbedSolution}, in the limit $x^+ \rightarrow +\infty$ we have 
\begin{equation}
 \Omega \sim e^{-2\phi} \sim -\lambda^2x^+ (x^--x^-_h)
 \,,
\end{equation}
so that the asymptotic behavior of $\rho$ is given simply by
\begin{equation}
 e^{-\rho} \rightarrow \bigg(1-\frac{x^-_h}{x^-}\biggr)^{1/2}
 = (1 + \lambda x^-_he^{\lambda \sigma^-})^{1/2}
 \,.
\end{equation}
Sending $P_O$ to $\mathcal{I}^+$ in \eqref{EQ_Sgen_General} gives
\begin{equation}
 S_\gen(I\cup R) = 2\Omega_Q 
 + 4\kappa\biggl\{\frac{\lambda}{2}(\sigma^+_Q - \sigma^-_Q)
 + \ln\lambda(\sigma^-_Q - \sigma^-_O)\biggr\}
 +4\kappa\ln\frac{\sigma^+_O}{\epsilon\sqrt{1+\lambda x^-_he^{\lambda\sigma^-_O}}}
\,,
\end{equation}
where the last term is IR and UV divergent but do not affect the extremization as it does not depend on the location of $P_Q$. We have recovered equation (3.6) of \cite{Hartman_Islands}. Before imposing the conditions $\partial_{\sigma^+_Q}S_\gen = 0 = \partial_{\sigma^-_Q}S_\gen$, let us express $\Omega$ in terms of $\sigma^\pm$,
\begin{equation}
 \Omega = e^{\lambda(\sigma^+-\sigma^-)} - \frac{\kappa}{2}\lambda(\sigma^+-\sigma^-) + \frac{m}{\lambda}\bigg(1-\frac{e^{\lambda\sigma^+}}{\lambda x^+_0}\bigg)\theta(x^+-x^+_0)
 \,,
\end{equation}
so that
\begin{align}
 S_\gen(I\cup R) & = 2e^{\lambda(\sigma^+_Q-\sigma^-_Q)}
 + \lambda\kappa(\sigma^+_Q - \sigma^-_Q)
 + 4\kappa\ln\lambda(\sigma^-_Q - \sigma^-_O)
 \nonumber
 \\
 & + 2\frac{m}{\lambda}\bigg(1-\frac{e^{\lambda\sigma^+_Q}}{\lambda x^+_0}\bigg)\theta(x^+_Q-x^+_0)
 +4 \kappa\ln\frac{\sigma^+_O}{\epsilon\sqrt{1+\lambda x^-_he^{\lambda\sigma^-_O}}}
 \,.
 \label{EQ_Boulware_Sgen}
\end{align}
Then we have
\begin{equation}
 \partial_{\sigma^+_Q}S_\gen = 2\lambda e^{\lambda\sigma^+_Q}\big[e^{-\lambda\sigma^-_Q} + \lambda x^-_h\theta(x^+_Q-x^+_0)\big] + \lambda\kappa
 = 0 \Leftrightarrow
 e^{-\lambda\sigma^+_Q} = -\frac{2}{\kappa}\big[e^{-\lambda\sigma^-_Q} + \lambda x^-_h\theta(x^+_Q-x^+_0)\big]
 \,,
\end{equation}
and
\begin{equation}
 \partial_{\sigma^-_Q}S_\gen = -2\lambda e^{\lambda(\sigma^+_Q-\sigma^-_Q)} - \lambda\kappa + \frac{4\kappa}{\sigma^-_Q - \sigma^-_O} = 0 \Leftrightarrow
 \frac{4\kappa}{\sigma^-_Q - \sigma^-_O}
 = \lambda\kappa + 2\lambda e^{\lambda(\sigma^+_Q-\sigma^-_Q)}
 \,.
\end{equation}
A quick analysis shows that there is no solution for $x^+_Q < x^+_0$, so that if there is an island, it can only be located after the shock $x^+_Q > x^+_0$. We can thus omit the Heaviside factor and combine these two equations to write them as
\begin{equation}
 -\frac{\lambda^2x^-_h}{2\kappa}(\sigma^-_Q-\sigma^-_O)
 = e^{-\lambda\sigma^+_Q} = -\frac{2}{\kappa}(e^{-\lambda\sigma^-_Q} + \lambda x^-_h)
 \,.
\end{equation}
Up to some constants which come due to differences in coordinate definitions, the above equation is the same as (3.7) of \cite{Hartman_Islands}. Expressing them in the $x^\pm$ coordinates, we get the following set of equations:
\begin{equation}
 \left\{
 \begin{aligned}
  & \lambda^2x^+_Q(x^-_Q - x^-_h) = \frac{\kappa}{2}
  \,, \\
  & x^-_O = x^-_Qe^{4\bigl(1-\frac{x^-_Q}{x^-_h}\bigr)}
  \equiv f(x^-_Q)
  \,, \\
  & x^+_Q > x^+_0
  \,.
 \end{aligned}
 \right.
 \label{EQ_Boulware_Island_System}
\end{equation}
The first equation tells us that the island, when it exists, lies on the curve obtained by reflecting the apparent horizon across the line $x^-=x^-_h$, as shown on figure \ref{FIG_BoulwareIsland}. This curve starts at the point $\big(x^+_0, x^-_h + \frac{\kappa}{2\lambda^2x^+_0}\big) = \big(x^+_0, \frac{1}{2\lambda^2x^+_0}(\kappa-2m/\lambda)\big)$ and is partially outside the singularity only if
\begin{equation}
 \frac{1}{2\lambda^2x^+_0}\bigg(\kappa-\frac{2m}{\lambda}\bigg)
 < -\frac{\kappa}{2\lambda^2x^+_0}
 \quad \textit{i.e.} \quad \frac{m}{\lambda} > \kappa
 \,.
\end{equation}
In the following we will work in the large $m$ limit, $\frac{m}{\lambda} >> \kappa$, so that this condition will always be satisfied. On the contrary, if the mass is too small, $m < \lambda\kappa$, then the island curve is behind the singularity and as such there is no quantum extremal surface. This suggests that one cannot derive a Page curve in that case.

\begin{figure}[htbp]
 \centering
 \includegraphics[scale=0.7]{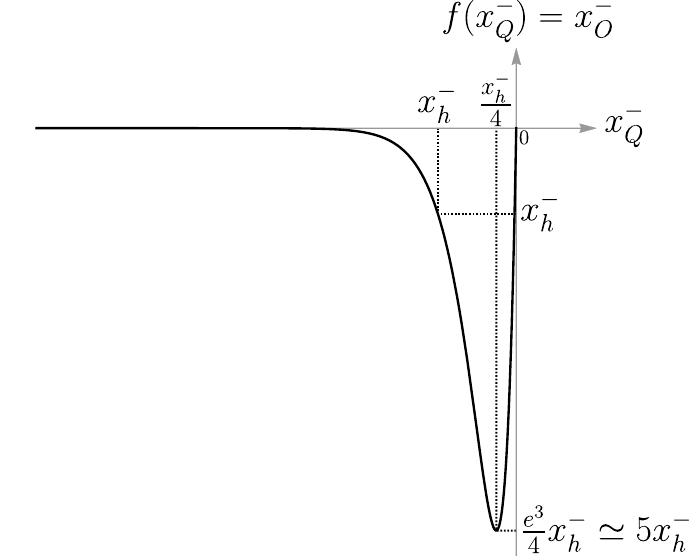}
 \caption{Plot of the function $f(x^-_Q) = x_O^-$. We see that there is no island for $x^-_O < 5x^-_h$ and that there are two possible islands for $x^-_O > 5x^-_h$.}
 \label{FIG_BoulwareIsland_Function}
\end{figure}

The condition $x^+_Q > x^+_0$ coupled to the first equation of \eqref{EQ_Boulware_Island_System} gives us bounds for $x^-_Q$ namely
\begin{equation}
 x^-_h < x^-_Q < x^-_h + \frac{\kappa}{2\lambda^2x^+_0}
 = \frac{1}{\lambda^2x^+_0}\biggl(\frac{\kappa}{2}-\frac{m}{\lambda}\biggr)
 \,.
 \label{EQ_BoundIsland}
\end{equation}
To find the location of the island as a function of the position of the observer $P_O$, one then has to invert the relation $x^-_O = f(x^-_Q)$ to find $x^-_Q$ as a function of $x^-_O$. The function $f(x^-_Q)$ is plotted on figure \ref{FIG_BoulwareIsland_Function}, it has a minimum at $x^-_Q = \frac{x^-_h}{4}$, with value $\frac{e^3}{4}x^-_h \simeq 5x^-_h$, which means that the island only exists for $x^-_O > \frac{e^3}{4}x^-_h$. As is clear from the graph, there are then two possible solutions for $x^-_Q$ corresponding to the two branches of the function $f$.
However one also has to keep in mind the condition $x^-_Q < x^-_h + \frac{\kappa}{2\lambda^2x^+_0}$ from \eqref{EQ_BoundIsland}. Indeed, if
\begin{equation}
 x^-_h + \frac{\kappa}{2\lambda^2x^+_0} < \frac{x^-_h}{4}
 \quad \textit{i.e.} \quad
 \frac{m}{\lambda} > \frac{2}{3}\kappa
 \,,
\end{equation}
then only the left branch is to be considered. But we have seen previously that the island only exists if $\frac{m}{\lambda} > \kappa$, so that this last condition is always satisfied.

Let us then work in the same limit as in \cite{Hartman_Islands}, namely $\frac{m}{\lambda} \gg \kappa$, and with $\lambda\tilde{\sigma}^-_O$ of order $\frac{m}{\lambda}$. We recall that $\lambda(x^-_O-x^-_h) = -e^{-\lambda\tilde{\sigma}^-_O}$. Note that in this limit $\lambda x^-_h = -\frac{m}{\lambda^2x^+_0}$ is also large. Then one can solve \eqref{EQ_Boulware_Island_System} perturbatively. First, use the first equation to rewrite the second as
\begin{equation}
 x^-_O = x^-_Q e^{-\frac{2\kappa}{\lambda^2x^-_hx^+_Q}}
 \,.
\end{equation}
Then, since $x^+_Q > x^+_0$, the argument of the exponential is small in the limit considered. We can therefore expand it and after some algebra it leads to
\begin{equation}
 \lambda x^+_Q = e^{\lambda\sigma^+_Q} = \frac{3\kappa}{2}e^{\lambda\tilde{\sigma}^-_O} + O\biggl(\frac{1}{\lambda x^-_h}\biggr)
 \,, \quad
 -\lambda x^-_Q = e^{-\lambda\sigma^-_Q} = -\lambda x^-_h + \frac{1}{3}e^{-\lambda\tilde{\sigma}^-_O} +
 O\biggl(\frac{e^{-2\lambda\tilde{\sigma}^-_O}}{\lambda x^-_h}\biggr)
 \,.
 \label{EQ_Boulware_Island_PerturbativeSolution}
\end{equation}
These two equations can be rewritten as
\begin{equation}
 \lambda\sigma^+_Q = \lambda\tilde{\sigma}^-_O
 + \ln\frac{3\kappa}{2}
 + O\biggl(\frac{e^{-\lambda\tilde{\sigma}^-_O}}{\lambda x^-_h}\biggr)
 \,, \quad
 -\lambda\sigma^-_Q = \ln(-\lambda x^-_h)
 - \frac{e^{-\lambda\tilde{\sigma}^-_O}}{3\lambda x^-_h}
 + O\biggl(\frac{e^{-2\lambda\tilde{\sigma}^-_O}}{(\lambda x^-_h)^2}\biggr)
 \,.
\end{equation}
We also have
\begin{equation}
 e^{-\lambda\sigma^-_O} = -\lambda x^-_h + e^{-\lambda\tilde{\sigma}^-_O}
 \,,
\end{equation}
which leads to
\begin{equation}
 \ln \lambda(\sigma^-_Q - \sigma^-_O) = -\lambda\tilde{\sigma}^-_O - \ln(-\lambda x^-_h) + \ln\frac{2}{3}
 + O\biggl(\frac{e^{-\lambda\tilde{\sigma}^-_O}}{\lambda x^-_h}\biggr)
 \,,
\end{equation}
and
\begin{equation}
 \ln\sqrt{1+\lambda x^-_he^{\lambda\sigma^-_O}} = 
 -\frac{1}{2}\lambda\tilde{\sigma}^-_O - \frac{1}{2}\ln(-\lambda x^-_h) + O\biggl(\frac{e^{-\lambda\tilde{\sigma}^-_O}}{\lambda x^-_h}\biggr)
 \,.
\end{equation}
Finally we can put everything together in \eqref{EQ_Boulware_Sgen} and get at leading order
\begin{equation}
 S_\gen(\tilde{\sigma}^-_O) \simeq \kappa\lambda(\tilde{\sigma}^-_{EP}-\tilde{\sigma}^-_O)
 + 4\kappa\ln\frac{\sigma^+_O}{\epsilon}
 \,,
\end{equation}
where we used the fact that $\lambda\tilde{\sigma}^-_{EP} \simeq \frac{2m}{\lambda\kappa}$. Corrections are of order $\ln(-\lambda x^-_h) \sim \ln\frac{m}{\lambda}$. Note that here $\sigma^+_O$ acts as an IR regulator. In the same limit, the entanglement entropy without island can be computed using \eqref{EQ_Boulware_Physical_DeltaS_TildeSigmaO} to get
\begin{equation}
 S_\ent(\tilde{\sigma}^-_O) \simeq 2\kappa\lambda\tilde{\sigma}^-_O + 4\kappa\ln\frac{\sigma^+_O}{\epsilon}
 \,,
\end{equation}
so that the island entropy is given by
\begin{equation}
 S_\isl(\tilde{\sigma}^-_O) = \kappa\lambda \min[2\tilde{\sigma}^-_O, \tilde{\sigma}^-_{EP} - \tilde{\sigma}^-_O]
 + 4\kappa\ln\frac{\sigma^+_O}{\epsilon}
 \,.
\end{equation}
This is once again in agreement with equation (3.17) of \cite{Hartman_Islands}. For $\tilde{\sigma}^-_O \leq \frac{1}{3}\tilde{\sigma}^-_{EP}$ this entropy is growing, and then it starts decreasing, exhibiting a Page curve behavior. Therefore, if the island procedure is to be trusted, it suggests that the evaporation of the black hole is indeed unitary. However we have seen that this does not seem to be the case if the mass of the shockwave is small enough, $m < \lambda\kappa$. Indeed in that case there is no island and we only get a growing contribution to the entropy.
\subsection{Island for the Unruh/Boulware state}
The asymptotic behavior at $\mathcal{I}^+$ is actually the same as in the Boulware case, so that when sending $P_O$ to infinity we once again get
\begin{equation}
 S_\gen(I\cup R) = 2\Omega_Q 
 + 4\kappa\biggl\{\frac{\lambda}{2}(\sigma^+_Q - \sigma^-_Q)
 + \ln(\sigma^-_Q - \sigma^-_O)\biggr\}
 +4\kappa\ln\frac{\sigma^+_O}{\epsilon\sqrt{1+\lambda x^-_he^{\lambda\sigma^-_O}}}
\,.
\end{equation}
The difference being that now $\Omega$ is given by
\begin{equation}
 \Omega = e^{\lambda(\sigma^+-\sigma^-)} - \frac{\kappa}{2}\lambda\sigma^+ + \frac{\kappa_2}{2}\lambda\sigma^- + \frac{M}{\lambda}
 + \frac{m}{\lambda}\biggl(1-\frac{e^{\lambda\sigma^+_Q}}{\lambda x^+_0}\biggr)\theta(x^+-x^+_0)
 \,,
\end{equation}
so that
\begin{align}
 S_\gen(I\cup R) & = 2e^{\lambda(\sigma^+_Q-\sigma^-_Q)}
 + \kappa\lambda\sigma^+_Q
 - (2\kappa_1 + \kappa_2)\lambda\sigma^-_Q
 + 4\kappa\ln\lambda(\sigma^-_Q - \sigma^-_O)
 + \frac{2M}{\lambda}
 \nonumber
 \\
 & + \frac{2m}{\lambda}\bigg(1-\frac{e^{\lambda\sigma^+_Q}}{\lambda x^+_0}\bigg)\theta(x^+_Q-x^+_0)
 + 4\kappa\ln\frac{\sigma^+_O}{\epsilon\sqrt{1+\lambda x^-_he^{\lambda\sigma^-_O}}}
 \,.
 \label{EQ_Unruh_Sgen}
\end{align}
It is interesting to note that one recovers the Boulware case by taking the limit $\kappa_1 \rightarrow 0$, so that $\kappa = \kappa_2$, and $M \rightarrow 0$. Extremizing the above equation, we obtain
\begin{equation}
 \partial_{\sigma^+_Q}S_\gen = 2\lambda e^{\lambda\sigma^+_Q}(e^{-\lambda\sigma^-_Q} + \lambda x^-_h\theta(x^+_Q-x^+_0)) + \lambda\kappa
 = 0 \Leftrightarrow
 e^{-\lambda\sigma^+_Q} = -\frac{2}{\kappa}\big[e^{-\lambda\sigma^-_Q} + \lambda x^-_h \theta(x^+_Q-x^+_0)\big]
 \,,
\end{equation}
and
\begin{equation}
 \partial_{\sigma^-_Q}S_\gen = -2\lambda e^{\lambda(\sigma^+_Q-\sigma^-_Q)} - \lambda(2\kappa_1 + \kappa_2) + \frac{4\kappa}{\sigma^-_Q - \sigma^-_O} = 0 \Leftrightarrow
 \frac{4\kappa}{\sigma^-_Q - \sigma^-_O}
 = \lambda(2\kappa_1 + \kappa_2) + 2\lambda e^{\lambda(\sigma^+_Q-\sigma^-_Q)}
 \,.
\end{equation}
Here we notice a first difference with respect to the Boulware solution, which is that there appears to be a solution before the shock ($x^+_Q < x^+_0$) given by
\begin{equation}
 \left\{
 \begin{aligned}
 & e^{\lambda(\sigma^+_Q - \sigma^-_Q)} = -\frac{\kappa}{2}
 \,, \\
 & \lambda(\sigma^-_Q - \sigma^-_O) = \frac{4\kappa}{\kappa_1}
 \,,
 \end{aligned}
 \right.
 \quad \textit{i.e.} \quad 
 \left\{
 \begin{aligned}
 & -\lambda^2x^+_Qx^-_Q = -\frac{\kappa}{2}
 \,, \\
 & x^-_Q = x^-_O e^{-\frac{4\kappa}{\kappa_1}}
 \,.
 \end{aligned}
 \right.
\end{equation}
However, the second equation of this system implies that $\sigma^-_Q < \sigma^-_O$, so that the points $P_O$ and $P_Q$ are timelike separated, which contradicts the initial hypothesis that the Cauchy surface $C$ should be everywhere spacelike or null (see figure \ref{FIG_Island_Slice}). Therefore there is actually still no solution before the shock.

\begin{figure}[htbp]
 \centering
 \includegraphics[scale=0.7]{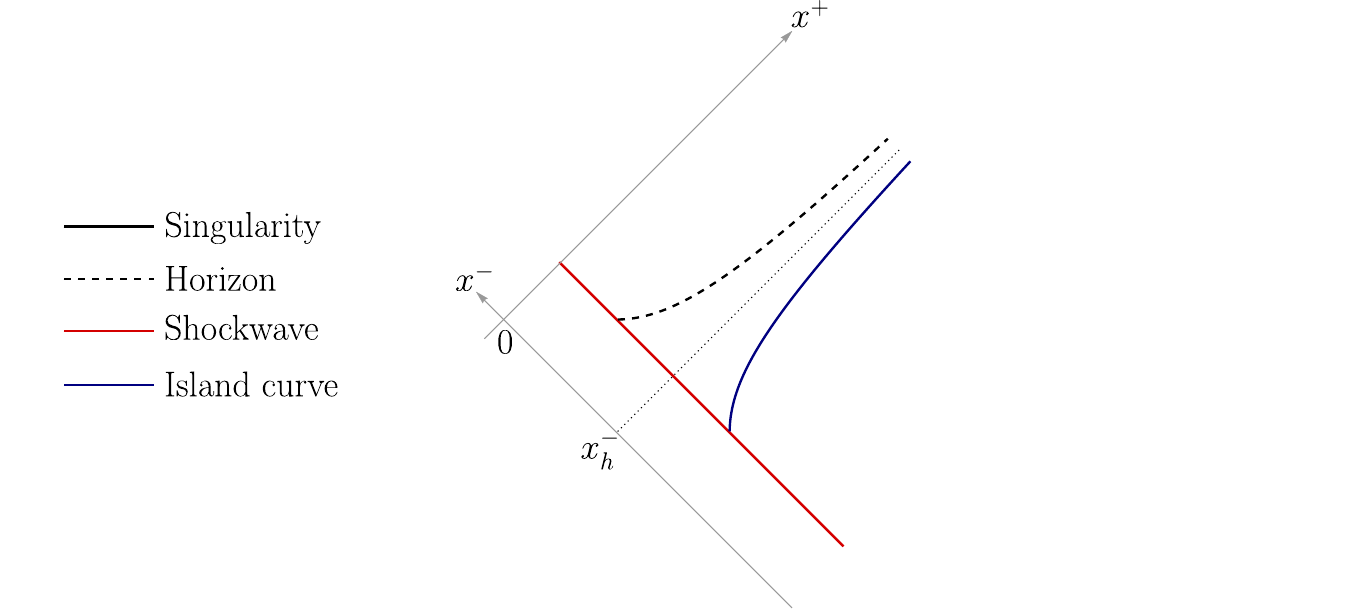}
 \caption{The island curve for the hybrid Unruh/Boulware state solution.}
 \label{FIG_UnruhIsland}
\end{figure}

After the shock ($x^+_Q > x^+_0$), we get
\begin{equation}
 \frac{-2\lambda^2x^-_h}{\frac{4\kappa}{\sigma^-_Q - \sigma^-_O}-\lambda\kappa_1} = e^{-\lambda\sigma^+_Q}
 = -\frac{2}{\kappa}(e^{-\lambda\sigma^-_Q} + \lambda x^-_h)\,,
\end{equation}
which gives
\begin{equation}
 \left\{
 \begin{aligned}
  & -\lambda^2 x^+_Q(x^-_Q - x^-_h) = -\frac{\kappa}{2}
  \,, \\
  & x^-_O = x^-_Q \exp \frac{4\kappa}{\kappa_1-\frac{\kappa x^-_h}{x^-_Q - x^-_h}} \equiv f(x^-_Q)
  = x^-_Q\exp\frac{4\kappa(x^-_Q-x^-_h)}{\kappa_1x^-_Q-(\kappa+\kappa_1)x^-_h}
  \,, \\
  & x^+ > x^+_Q
  \,.
 \end{aligned}
 \right.
 \label{EQ_Unruh_Island_System}
\end{equation}
The island is thus still located on the curve $-\lambda^2x^+(x^--x^-_h) = -\frac{\kappa}{2}$, which is also the reflection of the apparent horizon across the $x^-=x^-_h$ line, see figure \ref{FIG_UnruhIsland}. However, since $\kappa < 0$ here, the island is located outside of the apparent horizon.

%The function $f(x^-_Q)$ has a discontinuity at $x^-_Q = \Big(1+\frac{\kappa}{\kappa_1}\Big)x^-_h$, which is only relevant it is negative, that is if $\kappa_1>-\kappa$ \textit{i.e.} $2\kappa_1 > -\kappa_2$. Approaching this discontinuity from the left $f(x^-_Q)$ goes to $-\infty$ and it goes to $0$ approaching from the right.
%
%If $\kappa_1 = -\kappa$ then this discontinuity is at $x^-_Q = 0$.
%
%If $\kappa_1 < -\kappa$ then there is no physically relevant discontinuity and the function $f(x^-_Q)$ has a maximum and a minimum and satisfies
%\begin{equation}
% \lim_{x^-_Q \rightarrow -\infty}f(x^-_Q) = -\infty
% \,, \quad
% f(0) = 0
% \,.
%\end{equation}

Let us follow the same procedure as in the Boulware case, taking $\frac{m}{\lambda} \gg 1$ with $\lambda\tilde{\sigma}^-_O$ of order $\frac{m}{\lambda}$. The bounds $x^+_0 < x^+_Q < +\infty$, coupled to the condition $\lambda^2x^+_Q(x^-_Q-x^-_h) = \frac{\kappa}{2}$, imply that
\begin{equation}
 \lambda x^-_h + \frac{\kappa}{2\lambda x^+_0} < \lambda x^-_Q < \lambda x^-_h
 \,,
\end{equation}
so that for $\lambda x^-_h$ large, $\lambda x^-_Q$ is also large but has to be located in a relatively small interval of size $\frac{-\kappa}{2\lambda x^+_0}$. Therefore we can expect that in this limit, the function $f(x^-_Q)$ will be monotonous on this interval and thus that there will be only one solution to the equation $x^-_O = f(x^-_Q)$. We will see now that this is indeed the case.

The second equation of \eqref{EQ_Unruh_Island_System} can be rewritten as
\begin{equation}
 x^-_O = x^-_Q \exp \frac{4\kappa}{\kappa_1 - 2\lambda^2x^-_hx^+_Q}
 \,,
\end{equation}
and since the argument of the exponential is once again small, we can expand it and solve perturbatively to get
\begin{equation}
 \lambda x^+_Q = e^{\lambda\sigma^+_Q} = \frac{3\kappa}{2}e^{\lambda\tilde{\sigma}^-_O} + O\biggl(\frac{1}{\lambda x^-_h}\biggr)
 \,, \quad
 -\lambda x^-_Q = e^{-\lambda\sigma^-_Q} = -\lambda x^-_h + \frac{1}{3}e^{-\lambda\tilde{\sigma}^-_O} +
 O\biggl(\frac{e^{-2\lambda\tilde{\sigma}^-_O}}{\lambda x^-_h}\biggr)
 \,.
\end{equation}
This is actually exactly the same as in the Boulware case (see \eqref{EQ_Boulware_Island_PerturbativeSolution}), except that here $\kappa$ is negative. Therefore we can take all the perturbative expansions obtained in the previous section and plug them into \eqref{EQ_Unruh_Sgen}. Since the only difference between \eqref{EQ_Boulware_Sgen} and \eqref{EQ_Unruh_Sgen} is the coefficient in front of $\lambda\sigma^-_Q$, which gives a $\ln(-\lambda x^-_h)$ contribution, the result at order $\frac{m}{\lambda}$ is the same, that is
\begin{equation}
 S_\gen(\tilde{\sigma}^-_O) = \frac{2m}{\lambda} - \kappa\lambda\tilde{\sigma}^-_O
 \,.
\end{equation}
Hence the island entropy has the same expression as in the Boulware case
\begin{equation}
 S_\isl(\tilde{\sigma}^-_O) \simeq
 \min\biggl[2\kappa\lambda\tilde{\sigma}^-_O, \frac{2m}{\lambda} - \kappa\lambda\tilde{\sigma}^-_O\biggr]
 + 4\kappa\ln\frac{\sigma^+_O}{\epsilon}
 \,.
\end{equation}
However, this time $\kappa$ is negative, which means that the growing contribution and the decreasing contribution exchange roles. The decreasing contribution (obtained without an island) is always negative (since $\tilde{\sigma}^-_O > 0$), and the growing contribution (obtained with the island) is always positive. Therefore we don't have a Page curve but rather a steadily decreasing curve, which corresponds to the entanglement entropy curve obtained previously.

These results seem to indicate that the inclusion of islands is not required for the hybrid solutions as per the prescription \eqref{EQ_SIslandDefinition}, and that one can simply study the leading entanglement entropy piece.

%%%%%%%%%%%%%%%%%%%%%%%%%%%%%%%%%%%%%%%%
\section{Conclusion}\label{sec:conclude}
%%%%%%%%%%%%%%%%%%%%%%%%%%%%%%%%%%%%%%%%

Let us finish with some short concluding remarks. First, we can say that the results obtained by \cite{Hartman_Islands,Gautason:2020} for the solution describing physical fields in the Boulware state are very interesting, as they provide an example where the black hole evolution appears to be unitary only if one consider the presence of islands. On the contrary, we were able to find a Page curve for the entanglement entropy of the radiation without islands for our hybrid solution. This suggests that when non-physical fields are present, the island procedure might not be necessary. Of course this is only a particular instance, and one may wonder whether this would also be the case in other theories of semiclassical gravity and also in four-dimensions, which is far more technically challenging. Some advancements were made in \cite{Germani:2015tda,Berthiere_2017} in three and four dimensions respectively, but they deserve further attention, especially in the context of islands.

It must also be noted that our hybrid solution is rather peculiar. Indeed, before the shockwave, the spacetime is not Minkowski but rather a black hole mimicker as we saw in \cite{Potaux_3}. Another point is that the presence of non-physical fields leads to a negative outgoing energy flux at future null infinity, and it appears to be a key ingredient to the presence of a Page curve. 
According to the interpretation suggested in \cite{Potaux_2}, the non-physical fields represent the partners of thermal particles created during the Hawking radiation process. These partner particles initially move inward, crossing the event horizon, but are theorized to later re-emerge, carrying the missing information. Whether this mechanism is universal and holds in higher-dimensional scenarios remains an open question.

%The physical meaning behind this is rather unclear at this point.
%Therefore it remains unclear how such an object would be formed in a physical process.

%Of course it would be interesting to analyze how these aspects translate to other theories of semi-classical gravity, the main goal being the four-dimensional case, which is far more technically challenging. Some advancements were made in \cite{Germani:2015tda,Berthiere_2017} in three and four dimensions respectively, but they deserve further attention, especially in the context of islands.

%%%%%%%%%%%%%%%%%%%%%%%%%%%%%
%\bigskip{}
%\hfill{}\rule[0.3ex]{0.6\columnwidth}{1pt}\hfill{}
%\goodbreak
\vspace{0.5 cm}
%\centerline{\bf Acknowledgements}
%\noindent
%%%%%%%%%%%%%%%%%%%%%%%%%%%%%%

\noindent {\bf Acknowledgements:} The work of DS is supported by the DST-FIST grant number SR/FST/PSI-225/2016, SERB MATRICS grant MTR/2021/000168 and SERB CRG grant CRG/2023/000904.

\newpage
\bibliographystyle{unsrt}
\bibliography{biblio}

\end{document}